# Strain-Enhanced Coherence in Curved hBN Quantum Emitters

*Eyal Shoham*[§,1], *Sukanta Nandi*[§,1], *Ayelet Teitelboim*[2], *Jeny Jose*[1], *Gil Atar*[2], *Ashwin Ramasubramaniam*[3,4], *Tomer Lewi*[1], *and Doron Naveh*[*,1]

[1] Faculty of Engineering and Bar-Ilan Institute for Nanotechnology and Advanced Materials, Bar-Ilan University, Ramat Gan 52900, Israel.

[2] Applied Physics Division, Soreq NRC, Yavne 81800, Israel.

[3] Department of Mechanical and Industrial Engineering, University of Massachusetts Amherst, Amherst, Massachusetts 01003, USA.

[4] Materials Science and Engineering Graduate Program, University of Massachusetts Amherst, Amherst, Massachusetts 01003, USA.

[*] E-mail: doron.naveh@biu.ac.il

## Abstract

Hexagonal boron nitride (hBN) hosts robust room-temperature single-photon emitters, yet their coherence is typically limited by phonon-induced dephasing and spectral broadening. Here, we show that thermally induced curvature in bulk-like hBN flakes provides a strain-enabled route to suppress defect–phonon coupling under ambient conditions. Nanoscale bubbles formed by thermal processing generate strong through-thickness strain gradients, which we directly probe by infrared nano-spectroscopy. These measurements reveal strain-induced splitting of in-plane phonon modes, evidencing a substantial local modification of the phonon density of states. Quantum emitters localized within these curved regions exhibit markedly enhanced room-temperature spectral purity, with Debye–Waller factors of 91% and narrower linewidths than emitters in flat regions. Photon correlation measurements confirm high-purity single-photon emission at room temperature. Supported by first-principles calculations, we attribute this behavior to strain-driven phonon redistribution, which depletes phonons in tensile regions and accumulates them in compressive regions, thereby creating locally phonon-suppressed environments for defect emitters. These results establish strain engineering as an effective route for phonon control in hBN and open a pathway toward high-coherence, room-temperature quantum light sources for integrated nanophotonic platforms.

## Introduction

Hexagonal boron nitride (hBN), a layered insulator with a bandgap of ~6 eV, has attracted considerable interest in the realm of 2D nanophotonics. [1,2] hBN is highly stabile under ambient conditions[3] and displays a range of noteworthy intrinsic properties, [3] including piezoelectricity, [4] hyperbolic dispersion in the mid-infrared, [5] strong second-order nonlinearities, [6,7] and the ability to host optically active defects at room temperature. [8,9] These characteristics have enabled a range of photonic phenomena such as ultra-confined tunable low-loss surface phonon polaritons, [10,11] super-resolution imaging, [12] and room-temperature single-photon emission.[8,13] Among all of these phenomena, single-photon emission in hBN is particularly compelling due to its spectral tunability from UV to near-infrared, [14,15] and its high photon purity, [14,16] making it a strong candidate for realizing room-temperature, on-chip quantum photonic devices. [17] Single photon emission in hBN typically originates from color centers, which are associated with crystallographic point defects or vacancies, [18,19] as well as extended structural deformations and elastic strain. [20,21] Strain is a control knob for the spectrum of photon emission from hBN SPEs.[22,23] Electron-beam induced wrinkle-assisted bubbles in hBN form a reproducible, mechanically stable emitters that benefit from up to threefold brightness stemming from an optical interference effect [24].

In this work (schematized in Figure 1), we study thick, bulk-like hBN flakes and report on the formation and optical response of thermally-induced strain gradients manifested in curved hBN "*nano-bubbles*" (NBs).[25,26] We rationalize this seemingly non-intuitive behavior with a model that links the spatial distribution of strain fields to the local phonon density of states, giving rise to an effective strain-induced redistribution of phonons in regions of tensile and compressive strain.

Single-photon emission from hBN color centers typically benefits from cryogenic cooling, which suppresses decoherence and reduces interactions between acoustic phonons and excited electrons, a process that contribute to photoluminescence broadening, as well as optical phonon scattering responsible for spectral sidebands. This enables narrow linewidths and high spectral purity. [25,26] Although hBN quantum emitters can operate at room temperature, [8] cryogenic cooling remains essential for improving photon coherence and indistinguishability, a key parameter for quantum information protocols, by reducing dephasing and stabilizing the emitter's charge and spin environment. [27] It also mitigates non-radiative recombination and background fluorescence, which improves signal-to-noise ratios and ensures a high degree of photon antibunching. [25,28] Our findings and phenomenological model suggest that an equivalent *non-cryogenic* process mediated by strain, can reduce phonon scattering and may offer a facile route towards high-performance SPEs in hBN at *room temperature*.

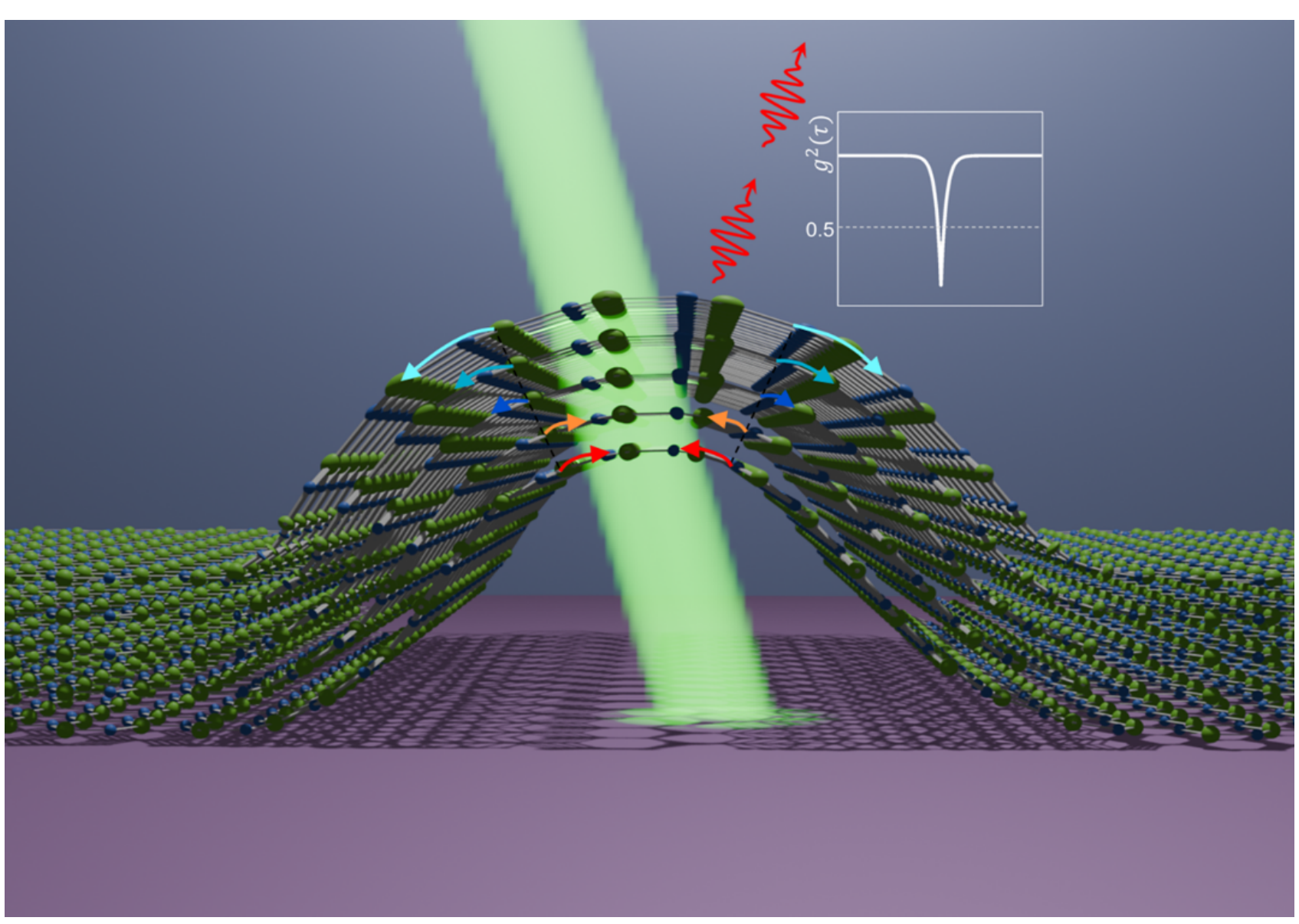

**Figure 1: Strain-Phonon Cooling.** Schematic cross-section of hBN nano-bubble with a through-thickness strain gradient across layers, leading to depletion of phonons from the top layers (tensile strain) and accumulation at the bottom layers (compressive strain). Upon excitation, single photons with high purity are emitted from color centers in the *"cooled"* phonon-depleted regions.

## Results and Discussion

Hexagonal boron nitride (hBN) flakes were mechanically exfoliated onto a 90 nm $Si/SiO_2$ substrate and subjected to rapid thermal annealing at 750 °C in ambient air (see Methods). Following annealing, nanoscale bubbles were identified near the edges of the hBN flakes. A photoluminescence (PL) intensity map is presented in Figure 2a, revealing localized bright spots close to the flake edges (inset of Figure 2a). The topography of one such NB, marked with a green circle in Figure 2a, was characterized using atomic force microscopy (AFM), as shown in Figure 2b. This NB appears as a hemispherical bubble at the flake's edge, with a thickness of ~45 nm (see Supplementary Figure 1). Infrared (IR) near-field spectra were collected using scattering-type scanning near-field optical microscopy (s-SNOM) at eleven points on the NB along line A and at one point labeled B, distant from the NB, as indicated in Figure 2b.

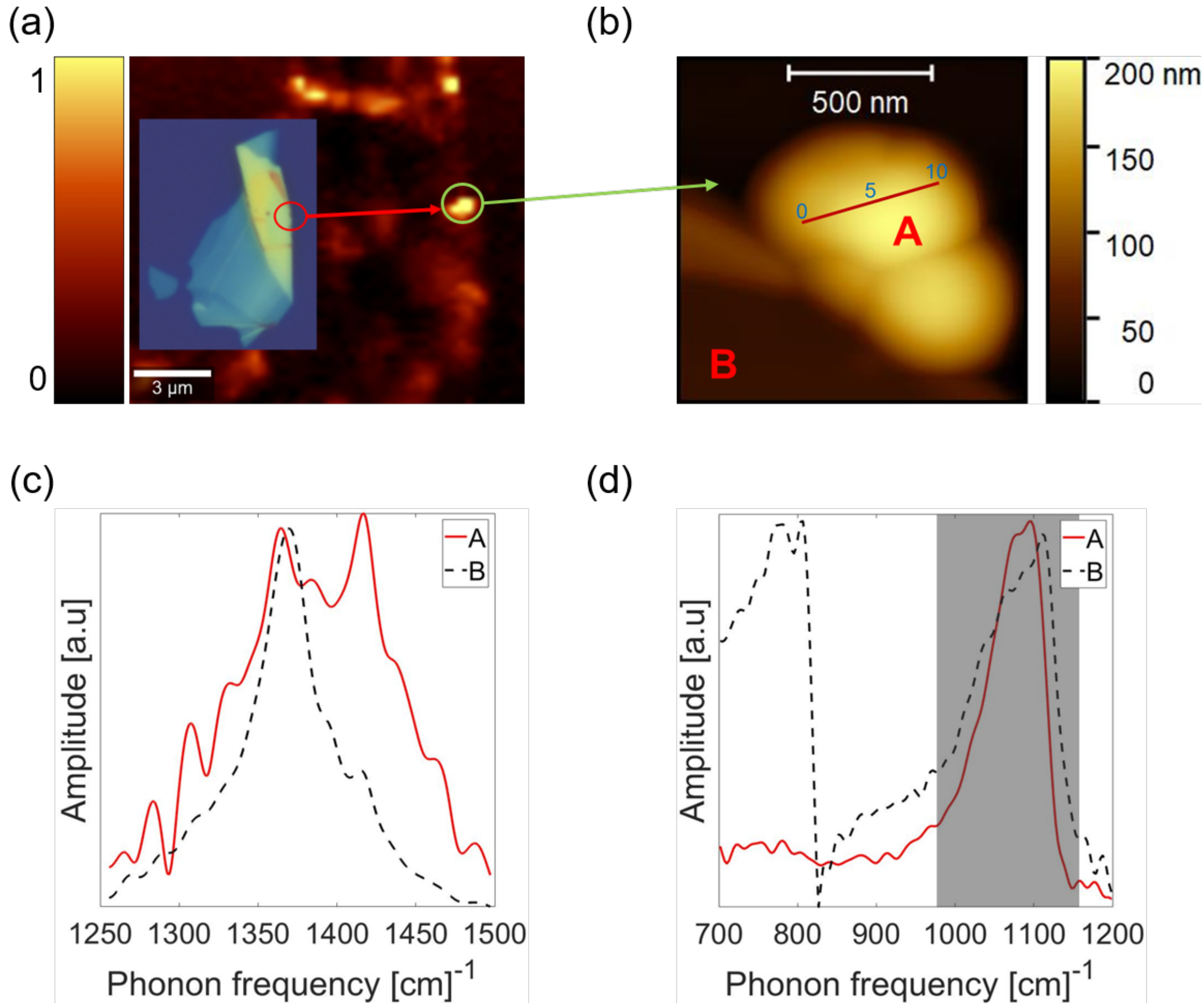

**Figure 2: Room temperature PL map and near-field infrared spectrum of hBN.** (a) PL Map of the flake with intensity representing the light collected for 660 (± 20) nm wavelength. The inset shows the bright-field image of the same flake. (b) AFM image of the single-photon emitter showing a corrugated structure (a NB) with maximum height of ~200 nm. (c, d) a comparison of phonon characteristics of region-A (hBN NB) and B (flat hBN region) using s-SNOM. The in-plane (c) and out-of-plane (d) phonon characteristics recorded at A and B show that the flat regions display typical phonon characteristics of hBN while the NB is affected by strain, as evidenced from the splitting of the in-plane phonon modes and complete suppression of the out-of-plane phonon at ~806 cm$^{-1}$. The grey region in (d) represents the $SiO_2$ phonon frequencies.

Infrared spectral analysis reveals that the transverse optical (TO) phonon modes recorded at the flat region (point B) exhibit a single, degenerate peak (Figure 2c; black dashed line). In contrast, spectra from point 5 on line A (NB center; red line) display clear mode splitting in the ~1283–1466 cm$^{-1}$ range, indicative of strain-induced lifting of phonon degeneracy. Furthermore, as shown in Figure 2d, the out-of-plane longitudinal optical (LO) phonon mode at ~806 cm⁻¹ is entirely suppressed in the NB, while the $SiO_2$ phonon at ~1113 cm⁻¹ is visible in both NB and flat regions.

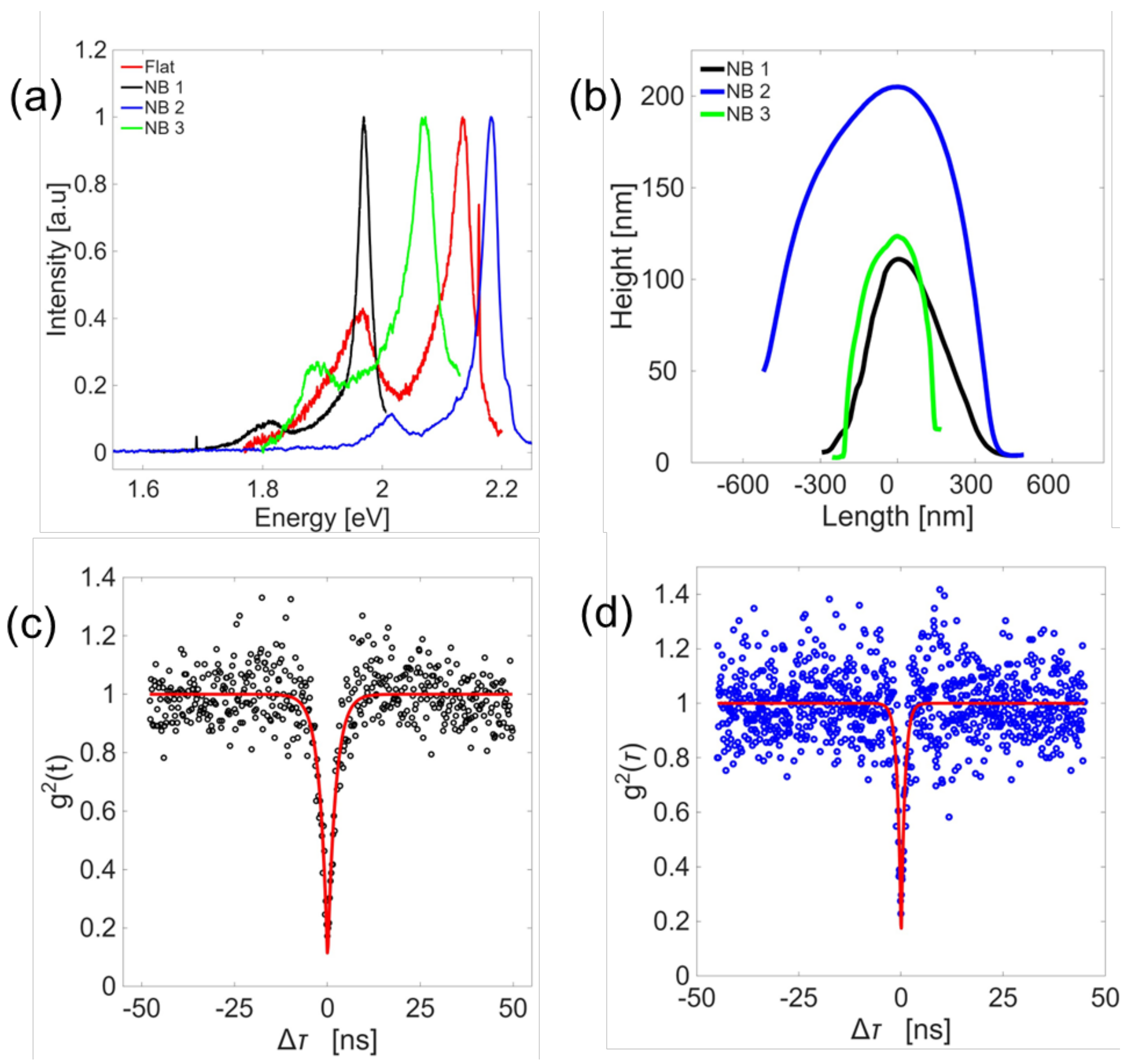

**Figure 3: Characteristics of quantum emission from hBN NB.** (a) PL spectra of hBN emitters on flat region (red), NB1 (black) and NB2 (blue), and NB3 (green). (b) The one-dimensional topographic profile of NBs. (c) the room temperature correlation measurement of the emitter on the hBN NB1, showing the emitter purity at room temperature $g^{(2)}(\tau = 0) = 0.09$ and a characteristic lifetime of $\tau_0 = 2.1$ ns. (d) the room temperature antibunching measurement of the NB2 emitter, showing the emitter purity at room temperature $g^{(2)}(\tau = 0) = 0.2$ and a characteristic lifetime of $\tau_0 = 0.9$ ns

Strain also affects optical transitions in hBN color centers strongly, as shown in Figure 3. Emission spectra from flat regions (red line) exhibits zero phonon lines (ZPLs) at 2.13 eV, while NB-based emitters display shifted emissions with ZPL at 1.97 eV (NB1 – black line), 2.17 eV (NB2 – blue line) and 2.07 eV (NB3 – green line). These shifts are consistent with $N_BV_N$-type (nitrogen anti-site nitrogen vacancy) defects induced by strain,[29,30] but could also arise from other point defects. The shifted PL of the NBs (Figure 3a) can also be associated to strain along the zigzag or armchair directions of hBN.[14,22]

The spectral width shown and the peak intensity of PSB to ZPL ratio in the emission of NB1 (black circles) and flat hBN emitter (red triangles) are displayed in Figures 4a-4b, suggest significantly diminished phonon scattering relative to a flat region. Photon correlation measurements (Figure 3c and Supporting Figure S3) from the room temperature emission from NBs confirm high-purity single-photon emission with $g^{(2)}(0)$ for all the emitters presented here (see Table 1).

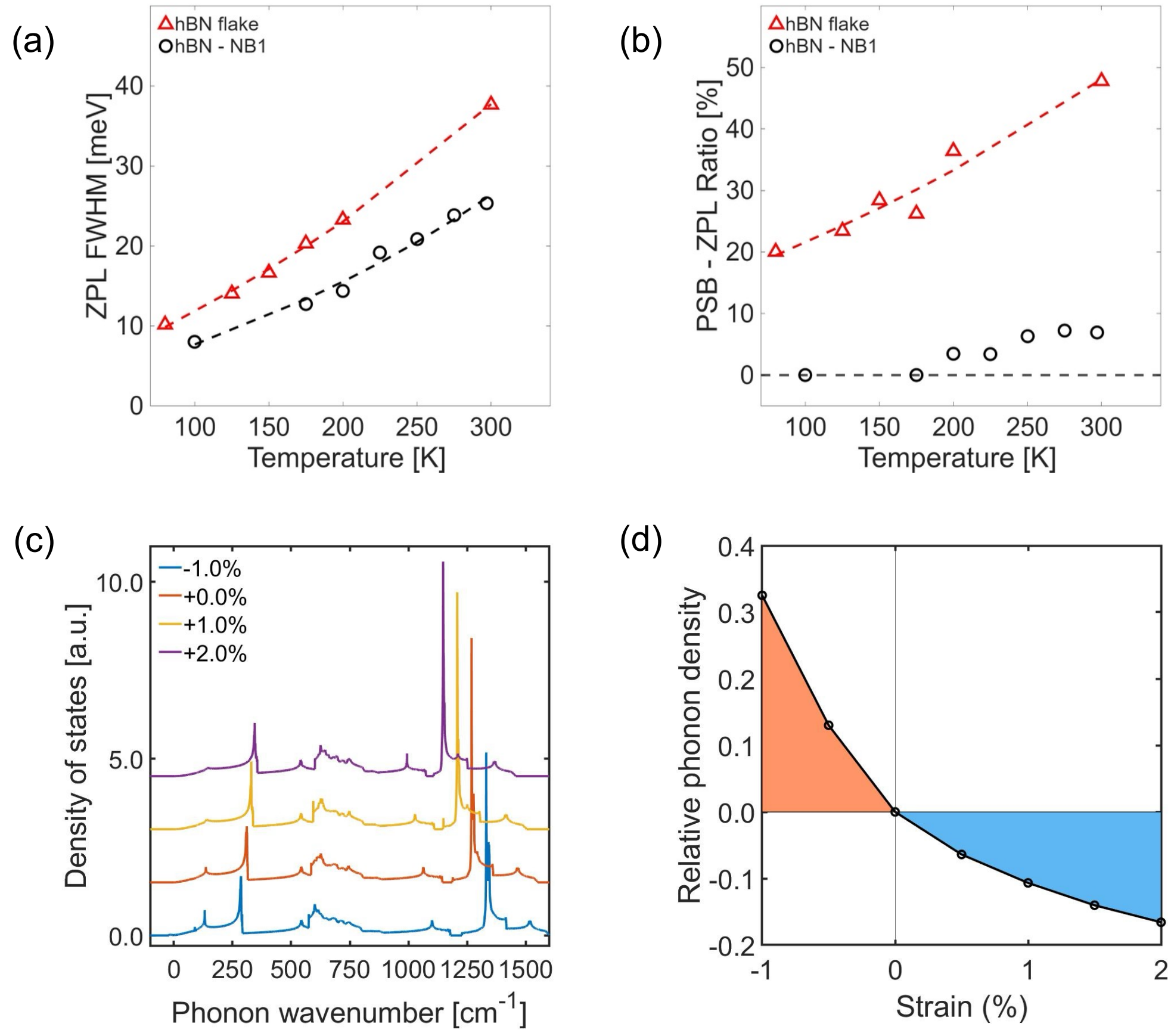

**Figure 4: Thermodynamic behavior.** (a) The temperature dependence of spectral width (FWHM) of PL emission from hBN color centers in the strain-free region compared to the room-temperature emission from NB1. (b) Ratio of the PSB-ZPL peak amplitudes in NB1 and flat hBN (c) Calculated phonon density of states (PDOS) at uniform in-plane strain in bulk hBN and (d) the normalized phonon accumulation or depletion relative to unstrained hBN, $\Delta n_{ph}(T_0, \epsilon)/n_{ph}(T_0, 0)$, as derived from the calculated PDOS and Eq. 1. The PDOS in (d) are shifted arbitrarily along the vertical direction for clarity.

A direct comparison of phonon scattering can be made with the Debye-Waller (DW) factor, defined as the integrated area ratio of the ZPL to PSB. The DW factor measures the degree of purity of the photon emission that is free from phonon scattered transitions. In addition, the Huang-Rhys (HR) model provides quantitative information on the coupling strength of radiative electronic transitions to phonons.[31,32] The larger the DW factor (and smaller HR), the weaker is the defect-phonon coupling. The key parameters of the room temperature measured NB emitters are listed in Table 1. A significant phonon decoupling in NBs is deduced from Table 1, where the room temperature DW factor in NBs is markedly high, reaching 91%.

**Table 1. Key parameters of quantum emitters measured at 300 K.**

| | **Thickness (nm)** | **Bubble diameter (nm)** | **Bubble height (nm)** | **Emission lifetime (ns)** | **ZPL energy (eV)** | **ZPL FWHM (meV)** | $g^{(2)}(0)$ | **DW (%)** | **HR** |
|---|---|---|---|---|---|---|---|---|---|
| Plane | 45 | - | - | 1.20 | 2.13 | 37.64 | $0.12 \pm 0.08$ | 59.79 | 0.5142 |
| NB1 | 38 | 380 | 120 | 2.19 | 2.07 | 39.7 | $0.09 \pm 0.01$ | 91.36 | 0.0904 |
| NB2 | 45 | 925 | 200 | 0.90 | 2.17 | 17.6 | $0.20 \pm 0.05$ | 85.54 | 0.1561 |
| NB3 | 32 | 500 | 104 | 2.15 | 1.97 | 24.8 | $0.29 \pm 0.06$ | 86.73 | 0.1423 |

We attribute the diminished phonon scattering in NBs to a strain-modified phonon density of states (PDOS), as shown in Figure 4c. At the measurement temperature, $T_0$=300 K, the strain-modified phonon population density is given by

$$n_{ph}(T_0, \epsilon) = \int_0^{\mathcal{E}_{max}} \frac{D(\mathcal{E}, \epsilon) d\mathcal{E}}{\exp\left(\frac{\mathcal{E}}{k_B T_0}\right) - 1}. \quad (1)$$

Here, $D(\mathcal{E}, \epsilon)$ denotes the strain-dependent PDOS calculated at 0 K from density functional theory, and $\mathcal{E}, \epsilon$ are the phonon energy and uniform in-plane strain, respectively. The strain-induced change in phonon population, $\Delta n_{ph}(T_0, \epsilon) = n_{ph}(T_0, \epsilon) - n_{ph}(T_0, 0)$, represents phonon redistribution under strain. As seen from Figure 4d, phonons accumulate in regions of compressive strain (orange) and are depleted in tensile regions (blue), driving the localized effective cooling of the quantum emitter in the latter case due to reduced phonon scattering. Within an hBN NB, the curved geometry near the apex (e.g., along line A in Figure 2b) causes a through-thickness strain gradient and lifts layer-wise phonon degeneracies, as observed from the shifted optical phonon mode (Figure 5a, 2c). The area under each deconvoluted peak of Figure 5a represents the relative thickness associated with the strain assigned to the phonon energy. [33]

**Table 2. Comparison of phonon scattering parameters in hBN single photon emitters.**

| *Ref* | *Preparation / platform* | *Temp. (K)* | *g²(0)* | *DW [%]* | *HR* | *ZPL [eV]* |
|---|---|---|---|---|---|---|
| *8* | Exfoliated hBN (multilayer); intrinsic defect | 77-300 | < 0.5 | 82 | 0.20 | 1.99 |
| *34* | AFM nano assembly with Au nanospheres | 300 | 0.24 | 75 | 0.29 | 2.145 |
| *35* | Mechanically exfoliated hBN defect ensemble | 300 | -- | 53 | 0.63 | 2.195 |
| *36* | Low-energy electron beam + annealing | 300 | ~0.25 | 50 | 0.6 | 2.145 |
| *37* | Arrayed isolated nanoflake on Au MW waveguide | 300 | 0.25 | 80 | 0.22 | 2.271 |
| *38* | Single-crystal hBN film; plasma-treated + hBN capping | 300 | 0.06 | 72.5 | 0.326 | 2.156 |
| *32* | AFM nanoindentation + annealing | 300 | 0.31 | 63.1 | 0.475 | 2.15 |
| *39* | Electron-beam irradiation | 4.5 | 0.505 | 50 | 0.69 | 2.265 |
| *40* | PLD-grown carbon-doped h-BN thin film | 300 | 0.015 | 45 | 0.79 | 2.136 |
| | Thermally induced nanobubble (**this work**) | 300 | 0.2 | 91.3 | 0.09 | 2.17 |

Table 2 benchmarks the performance of hBN quantum emitters across distinct preparation and defect-engineering approaches. Exfoliated multilayer hBN first established the material as an attractive host for weakly phonon-coupled quantum emission by exhibiting Debye–Waller factors up to 0.82.[8] Later strategies, including plasmonic nanoassembly,[34] low-energy electron-beam writing,[36] and engineered arrays of isolated single-spin defects, which also exhibit Debye–Waller factors approaching 0.8,[37] preserved high brightness and narrow-band emission while progressively improving emitter control and suppressing phonon coupling. In this landscape, our strain-engineered nano-bubble emitters define a particularly favorable regime, combining very large room-temperature Debye–Waller factors. These characteristics place them among the weakest phonon-coupled hBN emitters reported so far. Notably, the same behavior is reproduced for all nano-bubbles independent of size and shape.

The curvature of the bubble causes a through-thickness variation of the in-plane strain field, with maximum tension and compression being attained at the outermost and innermost layers, respectively. [41] Consequently, the local infrared spectra measured by the s-SNOM can be correlated with the through-thickness strain field distribution, as shown in Figure 5b, and with the continuous model presented in Figure 5c.

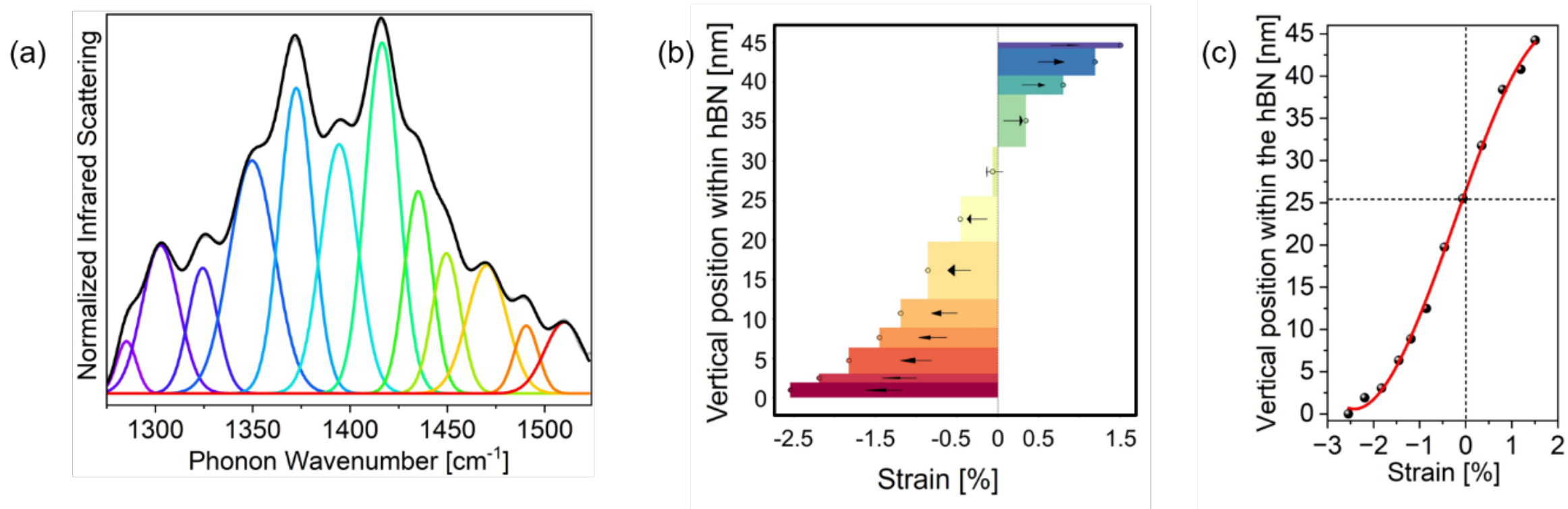

**Figure 5: Assessment of strain fields with s-SNOM infrared spectra of hBN NB.** (a) The IR spectrum and its deconvolution depicting strain distribution. (b) and (c) present the through-thickness strain distribution within the hBN NB.

According to the spectrally estimated strain distribution of Figure 5c and Figure S4, the peak-to-peak variation in the in-plane strain field can be up to 3.5% across the 32 and 45 nm thickness of the NB1 and NB2, respectively. This strong strain gradient drives the spatial redistribution of phonon density (Figure 4). Color centers located near the top layer at the apex of the hBN NB, where maximum tensile strain occurs, experience reduced phonon scattering compared to flat hBN (e.g., point B of Figure 2b). Conversely, color centers associated with point defects located near the bottom layer of the hBN NB, where compressive strain dominates, are expected to exhibit a higher rate of phonon scattering.

The optical response of localized quantum emitters in 2D materials is sensitive to strain with excitons being funneled towards/away from regions of tensile/compressive strain, and such curvature-induced strain effects are well studied in monolayers of several materials.[42–46]

However, the hBN films studied here are not only significantly thicker than monolayers but, more importantly, the optical transitions are between defect levels that are deep within the band gap such that there are no band-edge excitons involved. The effect of the tensile/compressive strains within a layer would, at most, lead to modulations of the transition energies at defect sites, i.e., an overall shift of the ZPL arising from local strain. This in of itself cannot explain the observed brightening of the SPEs. Consistently, the improved Debye–Waller and Huang–Rhys factors are not accompanied by systematic lifetime shortening across nanobubbles. We therefore attribute the dominant effect to strain-mediated suppression of defect–phonon coupling rather than cavity-enhanced spontaneous emission. A secondary contribution from partial substrate decoupling or mechanical isolation in the lifted nanobubble regions cannot be excluded,[47] particularly for the observed linewidth narrowing. However, the correlation between the local strain signatures observed by s-SNOM and the improved Debye–Waller and Huang–Rhys factors support strain-modified defect–phonon coupling as the dominant mechanism here.

## Conclusions

In conclusion, we show that thermally induced curvature in bulk-like hBN flakes creates a strain-engineered phonon landscape that enhances the coherence of defect-based quantum emission under ambient conditions. By combining infrared nano-spectroscopy, room-temperature single-photon measurements, and first-principles calculations, we establish a direct connection between through-thickness strain gradients, redistribution of the local phonon density of states, and reduced defect–phonon coupling in hBN nanobubbles. Emitters localized in these curved regions exhibit strongly improved room-temperature performance, including high Debye–Waller factors of up to 91%, reduced Huang–Rhys factors, narrowed linewidths, and robust antibunching, placing them among the weakest phonon-coupled hBN emitters reported to date. The fact that this behavior is reproduced across nanobubbles of different size and shape supports a materials-driven strain effect rather than a cavity-mediated mechanism. More broadly, these results identify strain-induced phonon redistribution as a practical route to emulate some of the benefits of cryogenic operation at room temperature, opening a pathway toward scalable, high-coherence quantum light sources for integrated nanophotonic and quantum-information platforms.

## Methods

**Sample preparation:** hBN was mechanically exfoliated from a commercial bulk crystal (HQ Graphene) on to an Si/$SiO_2$ chip (90 nm $SiO_2$). To increase the inhomogeneous strain created from the exfoliation, we induced a thermal shock that resulted in creation of NBs. Specifically, the chip was placed in a pre-heated oven for 10 minutes at 750°C, and left to cool, thereafter, in ambient air at room-temperature.

**PL spectra and mapping:** A 532 nm laser was employed for PL mapping of Figure 2a and to produce the spectra of Figs. 3a, 3b, and 4a at varying temperatures by using a commercial confocal microscope (Witec - alpha300 R). The map in Figure 2a shows bright luminescence from the nano-bubble (~2 times more intense) as compared to flat hBN regions, with sharp peaks for the spectra. Figure 2a shows an intensity map integrated between $660 \pm 25$ nm using a software band pass filter to depict the location of the emitter more clearly.

**Antibunching measurement:** Second order correlation measurements ($g^{(2)}(\tau)$) were performed on the NB (the strained regions) using a Hanbury Brown Twiss (HBT) setup. A 532 nm excitation laser was focused on a strained NB and its emission was collected through an objective (Zeiss x100 0.9 NA) passing through a dichroic mirror, which filters out the laser from the collected light. The resulting signal was directed either to a spectrometer (WITec VIS spectrometer) for spectral characterization or to an HBT for photon coincidence measurement. The HBT setup consists of a beam splitter that directs the light into two single-photon Avalanche Photodetectors (APDs; Micro Photon Devices, PD-100-CTE-FC) capable of about 35 ps temporal resolution. The APDs are connected to a single-photon counting module (Swabian instruments, Time-Tagger Ultra), that registers the signals for time-correlation analysis with less than 30 ps timing jitter. The time-tagger resolution was set to 50 ps and the measurement was performed over 15 minutes.

A single-photon sensitive CMOS camera KINETIX22 from Princeton instruments Teledyne was integrated with the system for wide-field inspection of the sample, aiming to image sample morphology and the dispersion of dilute emitters.

**Scanning Near-field Optical Microscopy (s-SNOM):** A commercial s-SNOM equipped with a nano-FTIR (Fourier Transform Infrared Spectroscopy) module (Neaspec/Attocube, Germany) was used for nanospectroscopy of the hBN samples. [48,49] The probe for the nano-FTIR measurements was a Pt/Ir-coated AFM tip oscillating at $\Omega \approx 233$ kHz with a tapping amplitude of ≈80 nm. This oscillating tip was illuminated by p-polarized mid-IR broadband radiation generated by a difference frequency generator (DFG) laser, covering a spectral range of ≈650–2200 $cm^{-1}$. The back-scattered signal from the samples was demodulated at higher harmonics of $\Omega$ and then detected using a liquid nitrogen cooled MCT (mercury cadmium telluride) detector. Interferograms were then acquired by measuring the demodulated detector signal at $3\Omega$ (exhibiting effective background suppression) as a function of the reference mirror position while keeping the tip position fixed in an asymmetric Michelson interferometer configuration. Fourier transformation of the interferogram signals then yielded complex-valued near-field spectra (amplitude and phase). Finally, the nano-FTIR spectrum was obtained by normalizing the measured spectrum of hBN with a reference spectrum recorded on silicon.

**Computational Modeling**: The phonon density of states was calculated using planewave density-functional theory (DFT), as implemented in the Vienna *Ab Initio* Simulation Package (VASP), [50,51] in conjunction with the phonopy package. [52,53] The projector-augmented wave method,[54,55] was used to treat core and valence electrons and the Perdew-Burke-Ernzerhof version of the generalized-gradient approximation was used to model electron exchange and correlation. [56] Dispersive interactions between hBN layers were modeled using the Tkatchenko-Scheffler method. [57] The *AA′*-stacked unit-cell of bulk hBN was relaxed using a kinetic energy cutoff of 700 eV, a $16 \times 16 \times 8$ *k*-point mesh, and a force (stress) tolerance of 0.001 eV/Å (0.001 eV/Å$^3$). Strained unit cells were modeled by applying equibiaxial strains to the *a*- and *b*-axis lattice vectors and selectively relaxing the *c*-axis vector (i.e., normal to the layers). The relaxed unit cells were used to construct $5 \times 5 \times 2$ supercells for calculating phonon modes using a finite-difference approach. As compressive in-plane strains lead to long-wavelength structural instabilities that cannot be captured using supercells that are tractable for DFT calculations, we do not consider strains below -1% by which point the structural instability, manifested by negative phonon modes - becomes apparent in the phonon dispersion and the phonon density of states (Figure 4b). Nevertheless, for the range of compressive strains considered here, the contribution from these negative modes to the phonon density of states is very small and is not expected to affect the overall conclusions.

**Acknowledgements**

DN, ES, AT and GA would like to thank the joint IAEC-UPBC Pazy Foundation for the support of this work under grant No. 256/20. AR gratefully acknowledges computing support from the University of Massachusetts Amherst. T.L. gratefully acknowledges support from the Israel Science Foundation (ISF), Grant No. 523/23, for this project

## Author Contributions

ES, AT, GA and DN designed and built experimental measurement setup for SPE characterization. ES and AT conducted measurements of SPEs. ES, JJ, AT, GA and DN prepared samples for measurements. TL and SN performed and analyzed s-SNOM measurements. AR and DN performed the theoretical analysis. All authors contributed to the final manuscript.

## Competing Interest

The authors declare no financial or competing interests.

# Supplementary Information

# Strain-Enhanced Coherence in Curved hBN Quantum Emitters

*Eyal Shoham*[§,1], *Sukanta Nandi*[§,1,2,3], *Ayelet Teitelboim*[4], *Jeny Jose*[1], *Gil Atar*[4], *Ashwin Ramasubramaniam*[5,6], *Tomer Lewi*[1], *and Doron Naveh*[*,1]

[1]*Faculty of Engineering and Bar-Ilan Institute for Nanotechnology and Advanced Materials, Bar-Ilan University, Ramat Gan 52900, Israel.*
[2]*Department of Electronics & Communication Engineering, Indraprastha Institute of Information Technology Delhi (IIIT-Delhi), New Delhi 110020, India*
[3]*Centre for Quantum Technology, Indraprastha Institute of Information Technology Delhi (IIIT-Delhi), New Delhi 110020, India*
[4]*Applied Physics Division, Soreq NRC, Yavne 81800, Israel.*
[5]*Department of Mechanical and Industrial Engineering, University of Massachusetts Amherst, Amherst, Massachusetts 01003, USA.*
[6]*Materials Science and Engineering Graduate Program, University of Massachusetts Amherst, Amherst, Massachusetts 01003, USA.*

[*]E-mail: doron.naveh@biu.ac.il

**Table of Contents**

·

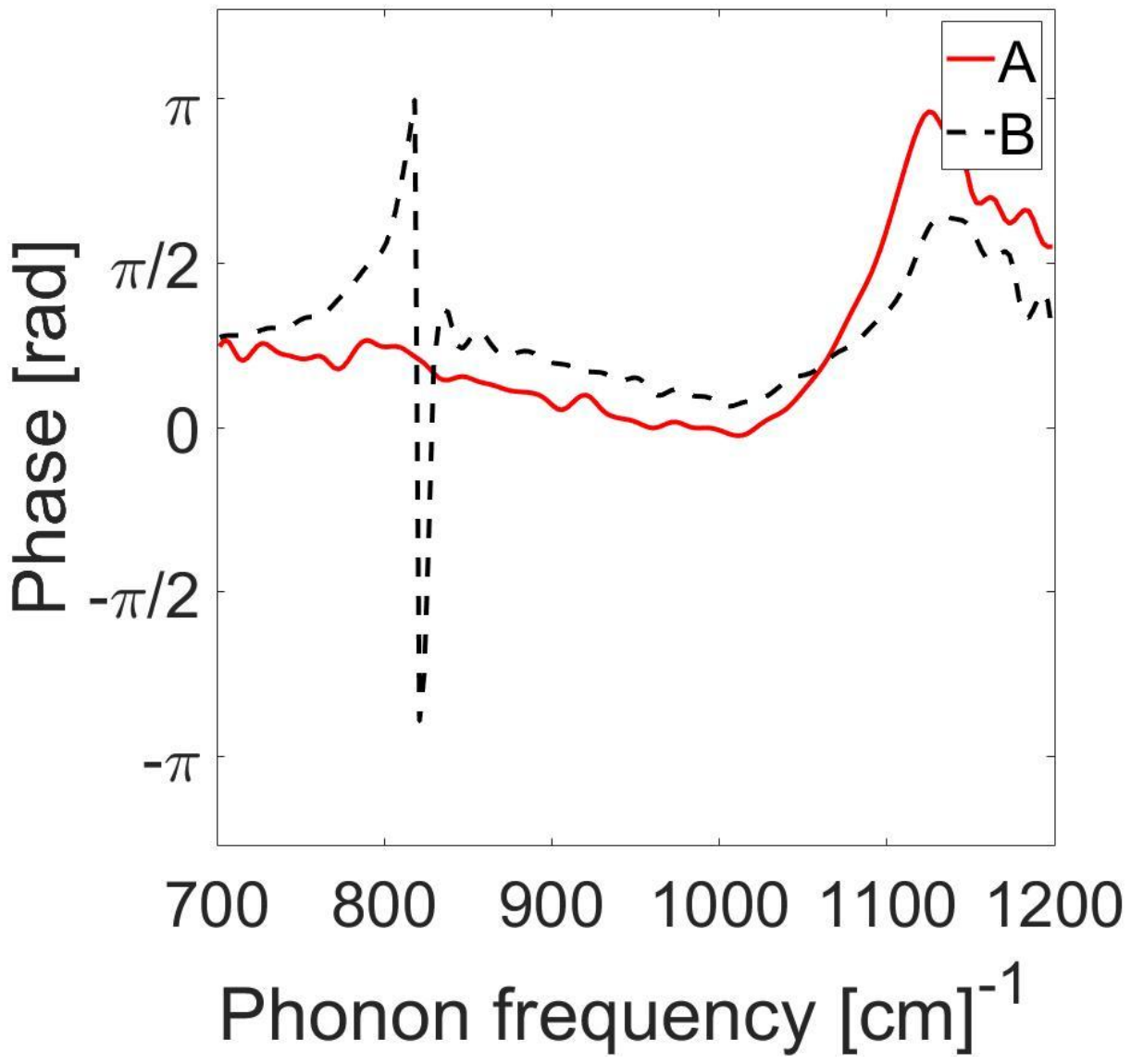

**Figure S1. s-SNOM optical phase for the out-of-plane LO phonon mode.** The corresponding phase spectra for Fig. 2d (main manuscript) showing complete absence of the out-of-plane phonon absorption (red solid line, A) on the hBN nano-bubbles NB2. The black dashed line (B) shows the out-of-plane phonon absorption from the hBN flake.

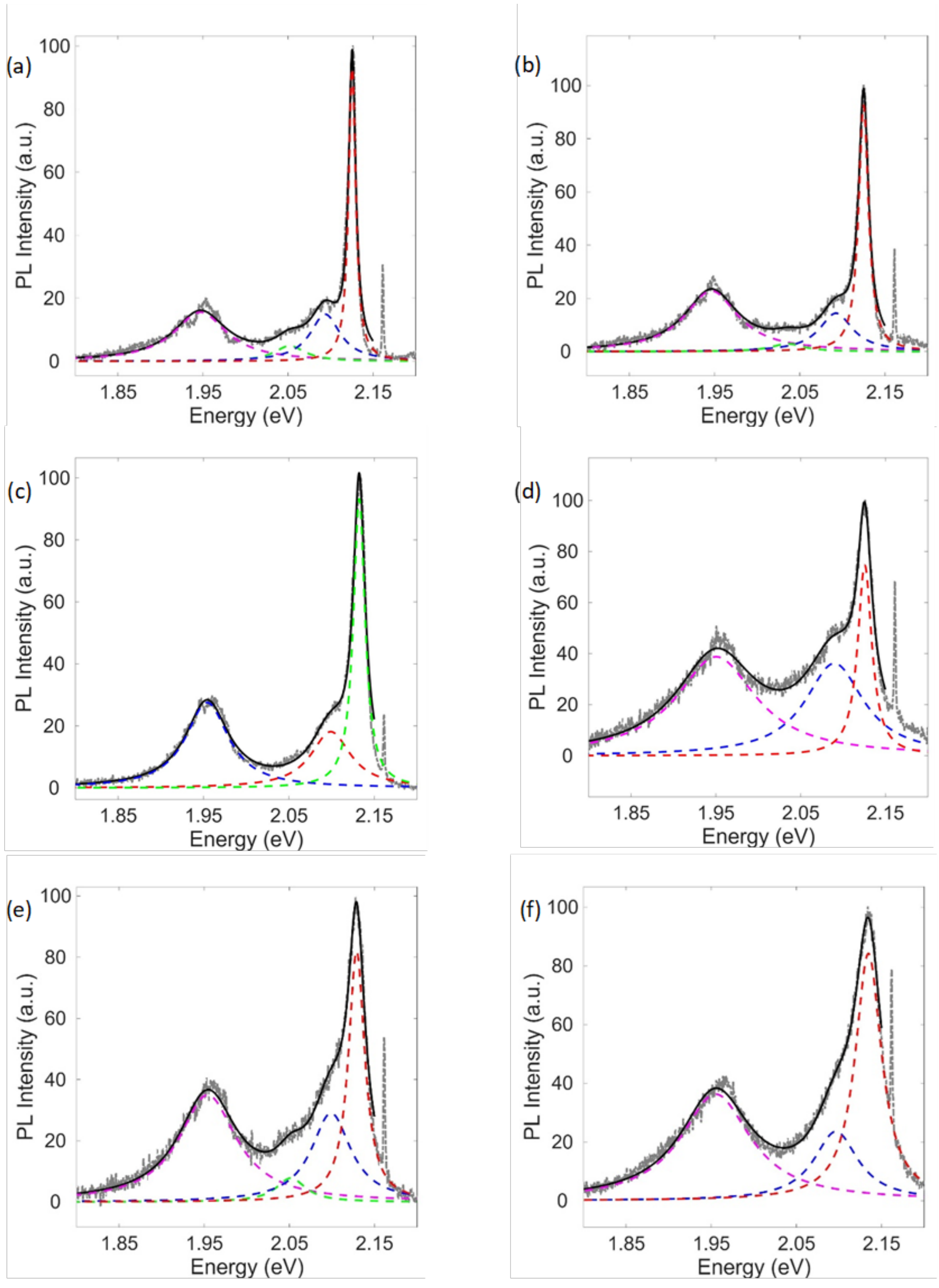

**Figure S2. PL spectra of hBN emitter at different Temperatures.** Deconvoluted spectra of emitter on the flat region of the same hBN flake measured at different temperatures: (a) 80K, (b) 125K, (c) 150K, (d) 175K, (e) 200K, (f) 300K with PSB/ZPL corresponding to Figure 4a, 4b in the main text.

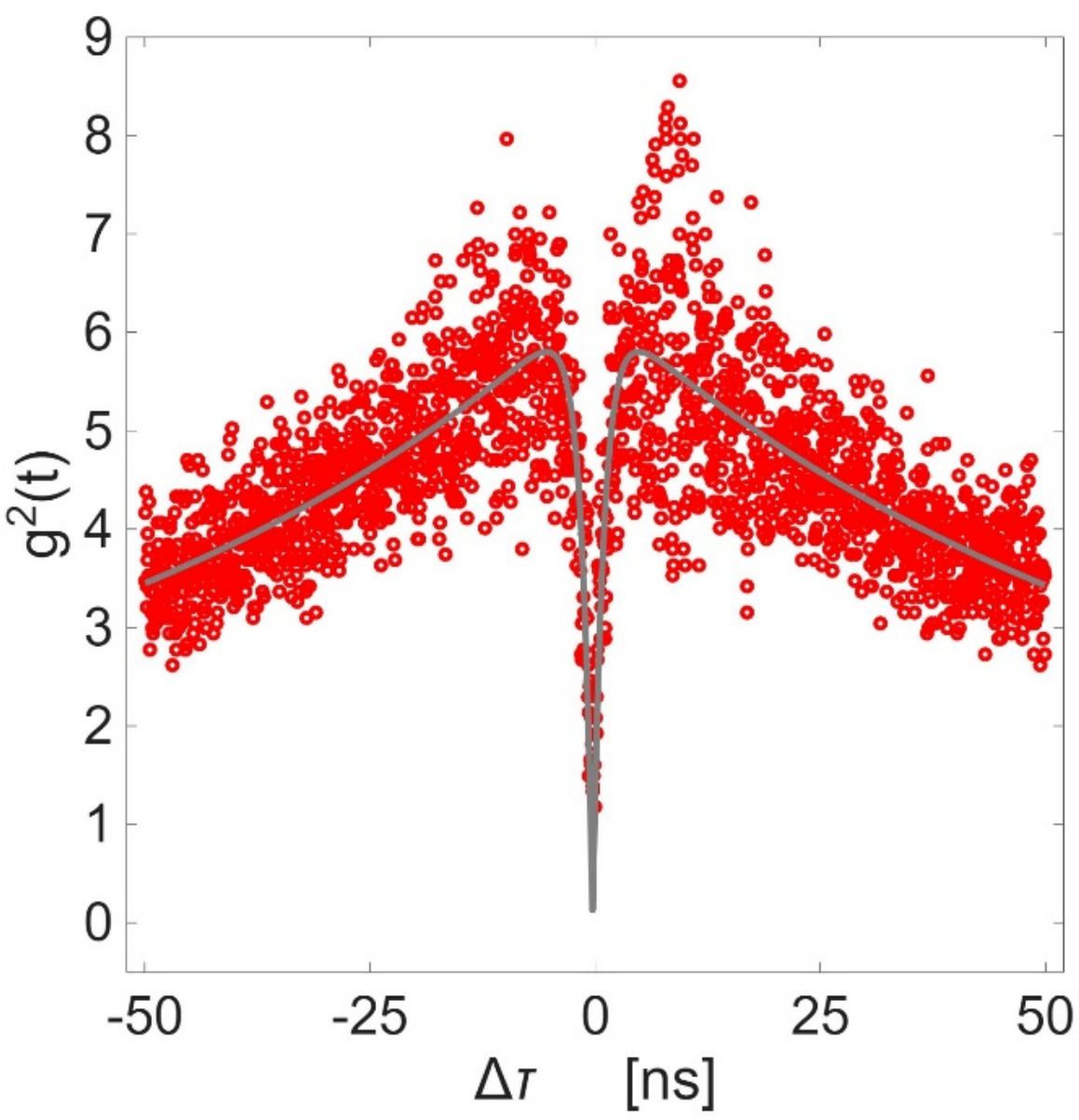

**Figure S3. Time correlation measurement of the hBN flake.** the room temperature antibunching measurement of the emitter on the flat region hBN, showing the emitter purity at room temperature $g^{(2)}(\tau = 0) = 0.12 \pm 0.08$ and a characteristic lifetime of $\tau_1 = 1.2$ ns and $\tau_2 = 64.5$ ns

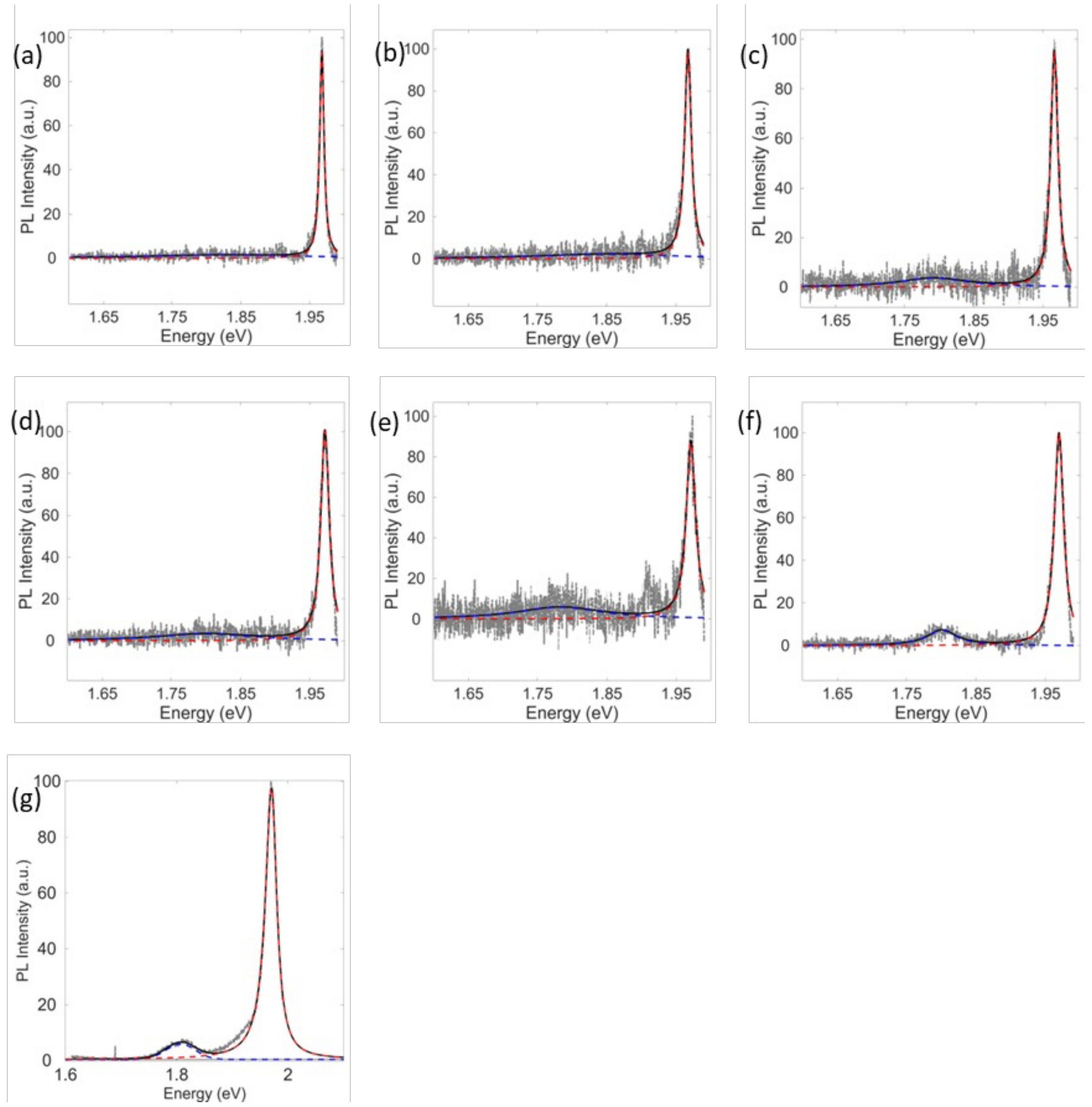

**Figure S4. PL spectra of hBN emitter NB1 at different Temperatures.** Deconvoluted spectra of emitter NB1 measured at different temperatures: (a-g) 100K, 175K, 200K, 225K, 250K, 275K, and 297K, respectively.

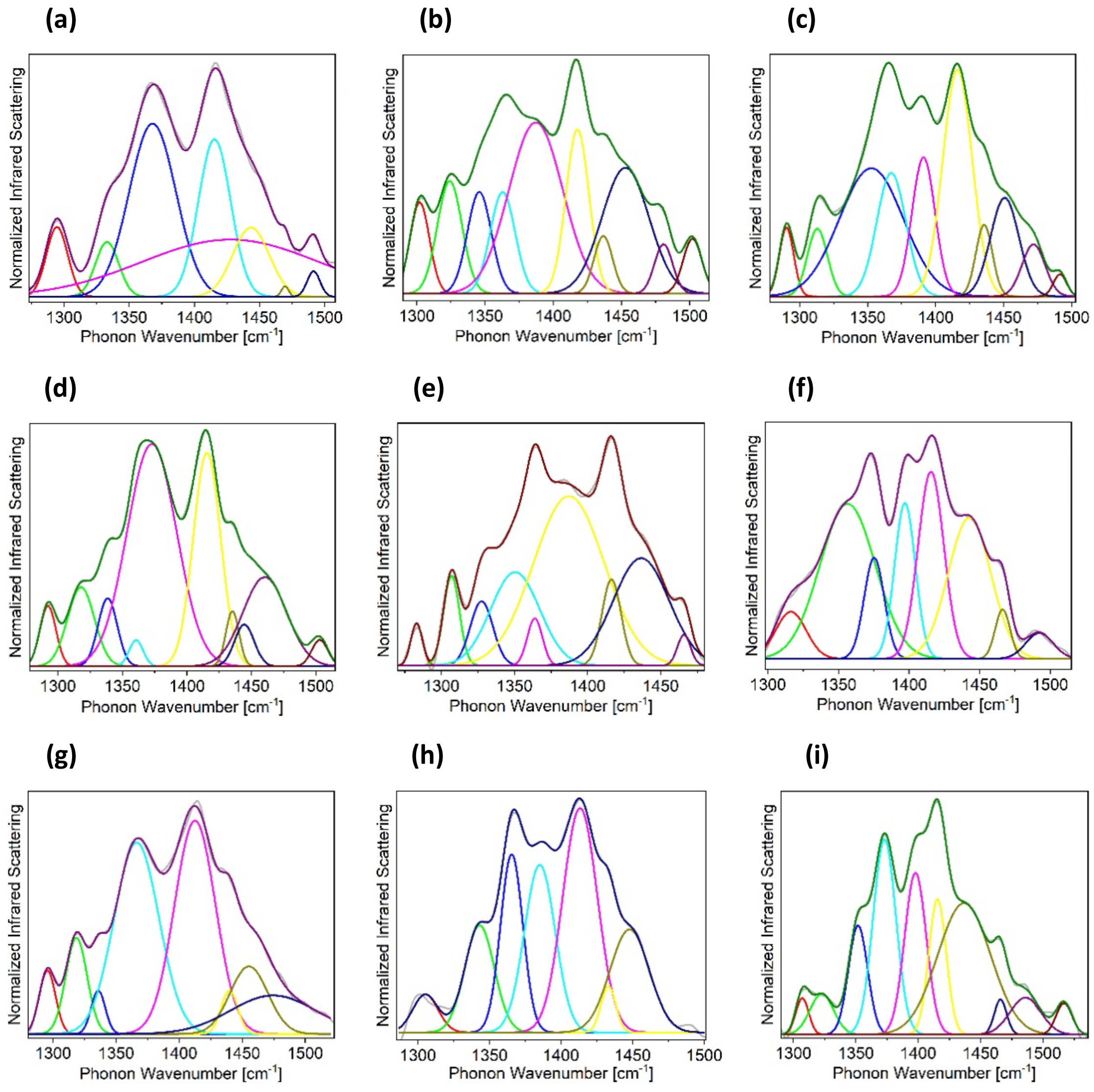

**Figure S5. Deconvoluted spectra of the in-plane phonon mode at different locations of the hBN NB2.** Locations are marked by the numbers 0-10 in correspondence to the positions of spectra acquisition along "line A" in Fig. 2b of the main manuscript are displayed from (a) to (i) and present the degeneracy lifted phonon characteristics as measured on the hBN nano-bubble.

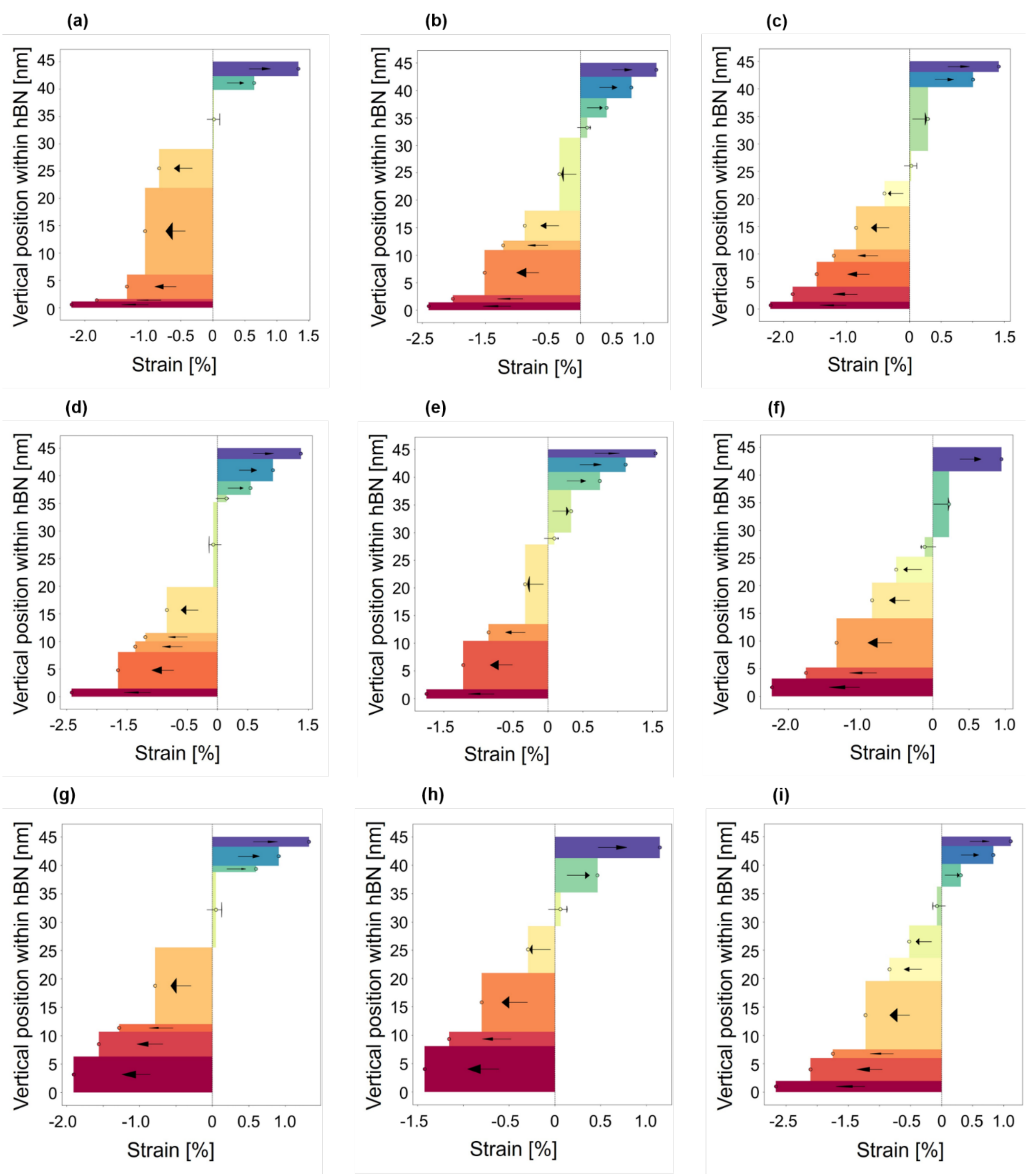

**Figure S6. The through-thickness strain distribution within the hBN NB. at different locations of the hBN NB2.** (a-i) represents the deconvoluted in-plane phonon spectrum across different locations on the hBN nano-bubble. These locations are marked by the numbers 0-10 while location number 4 is described in the main paper.

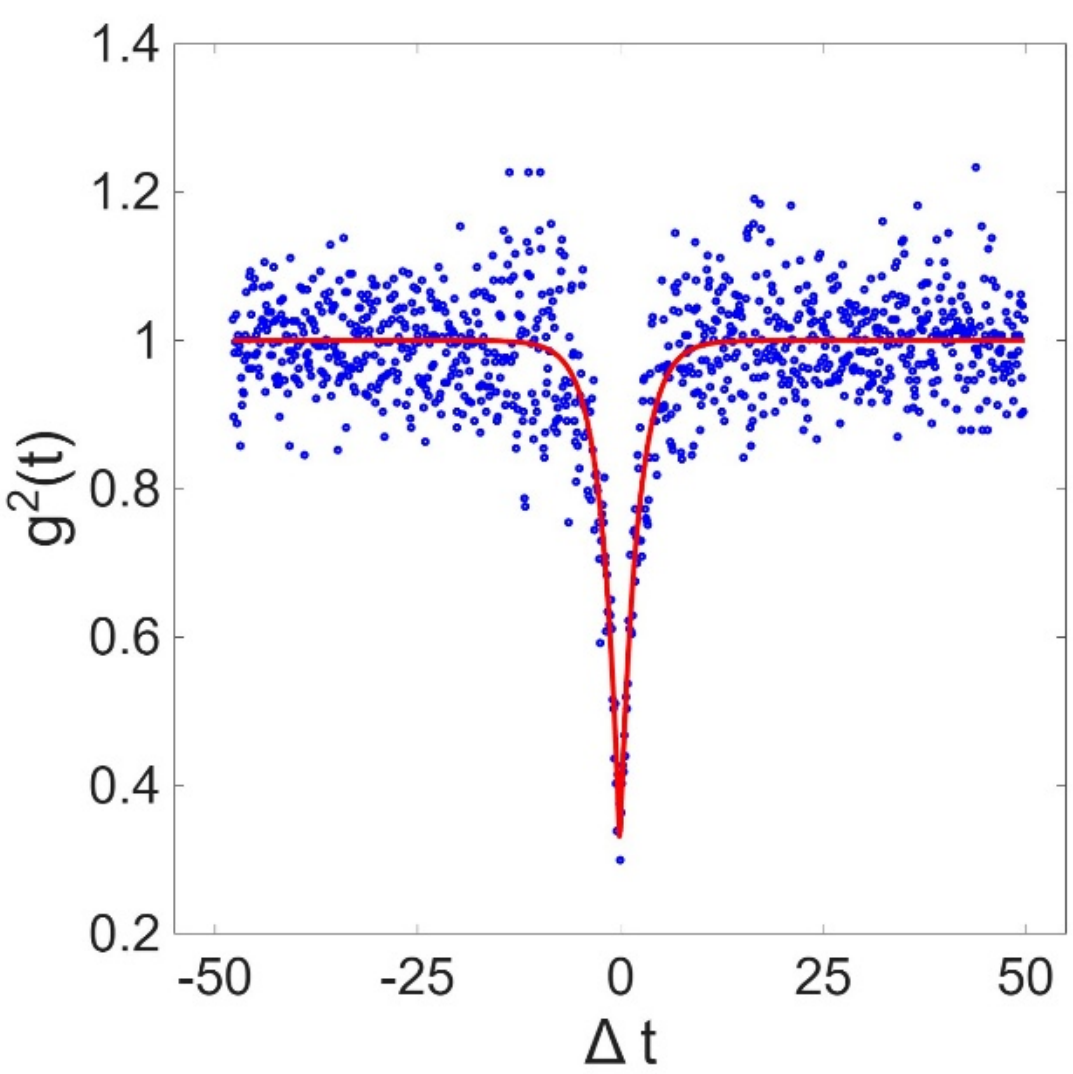

**Figure S7. Time correlation measurement of the hBN NB3.** the room temperature antibunching measurement of the emitter on the flat region hBN, showing the emitter purity at room temperature $g^{(2)}(\tau = 0) = 0.31 \pm 0.04$ and a characteristic lifetime of $\tau_0 = 2.19$ ns.

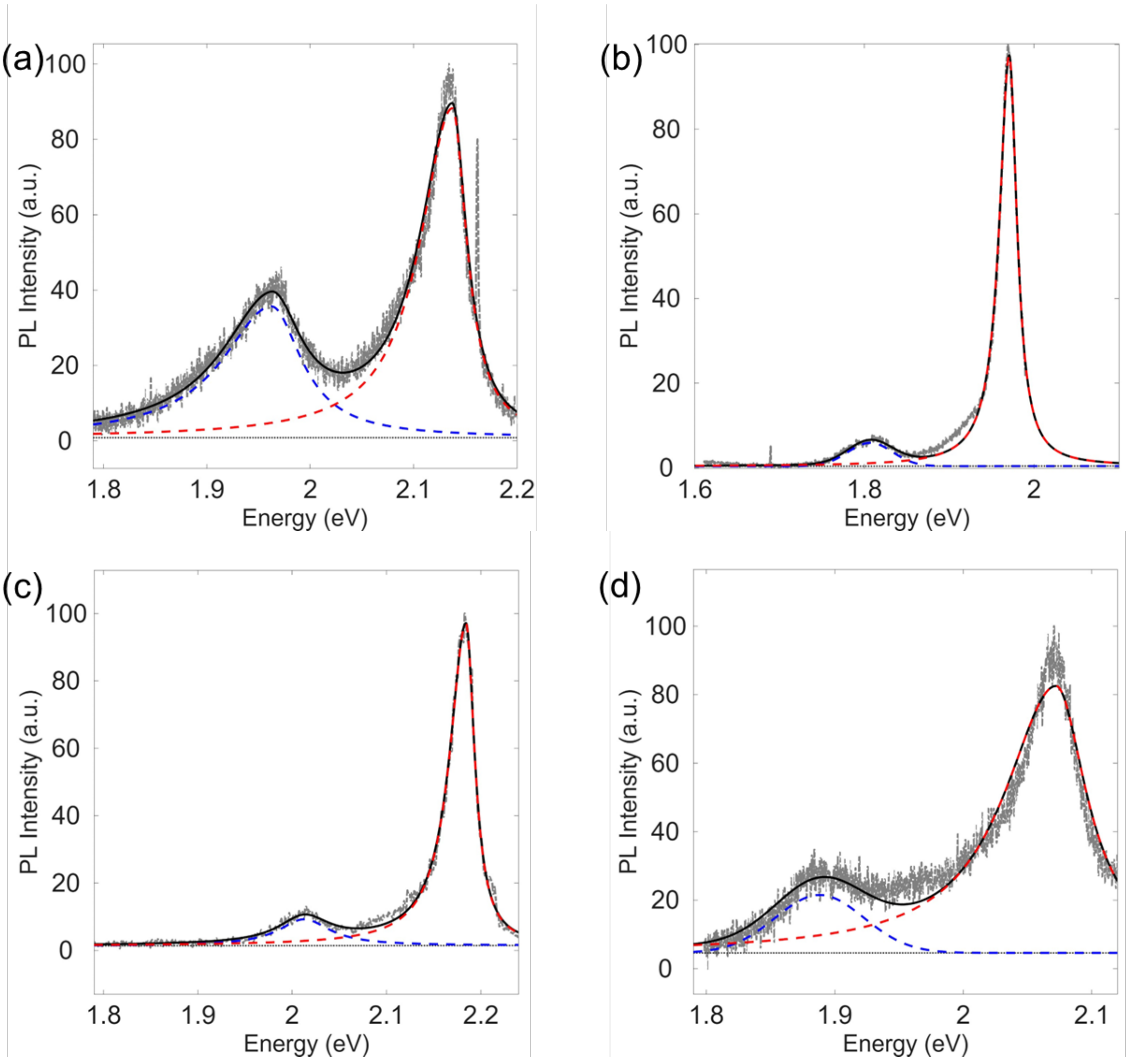

**Figure S8. Spectral fit for calculating the Debye-Waller (DW) and Huang-Rhys (HR) factors at RT.** The room temperature photoluminescence spectra (grey dots) of (a) flat hBN, NB1 (b), NB2 (c), and NB3 (d)fitted to ZPL (red dash), PSB (blue dash) is fitted to an asymmetric Pseudo-Voigt function. Black lines show the superposition of fitted peaks shown in dashes.

The DW factor was calculated by the area ratio of the fitted spectra with the equation $F_{(DW)} = \frac{I_{ZPL}}{I_{total}}$, where $I_{ZPL}$ is the areal integral of the ZPL peak and $I_{total}$ is the total area under the spectrum.

The HR factor was calculated by: $F_{(HR)} = -\ln\left(F_{(DW)}\right)$.